\documentclass[11pt,twoside]{article}
\usepackage{asp2014}

\aspSuppressVolSlug
\resetcounters

\begin{document}

\title{Amplitude modulation in the ZZ Ceti star GD 244}
\author{Zs.~Bogn\'ar,$^1$ M.~Papar\'o,$^1$ L.~Moln\'ar,$^{1,2}$ E.~Plachy,$^{1,2}$ and \'A.~S\'odor$^1$
\affil{$^1$Konkoly Observatory, MTA CSFK, Konkoly Th. M. \'ut 15-17, H-1121 Budapest, Hungary; \email{bognar.zsofia@csfk.mta.hu}}
\affil{$^2$Institute of Mathematics and Physics, Savaria Campus, University of West Hungary, 
K\'arolyi G\'asp\'ar t\'er 4, H-9700 Szombathely, Hungary}
}

\paperauthor{Zs.~Bogn\'ar}{bognar.zsofia@csfk.mta.hu}{}{Konkoly Observatory, MTA CSFK}{}{Budapest}{}{1121}{Hungary}
\paperauthor{M.~Papar\'o}{paparo@konkoly.hu}{}{Konkoly Observatory, MTA CSFK}{}{Budapest}{}{1121}{Hungary}
\paperauthor{L.~Moln\'ar}{lmolnar@konkoly.hu}{}{Konkoly Observatory, MTA CSFK}{}{Budapest}{}{1121}{Hungary}
\paperauthor{E.~Plachy}{eplachy@konkoly.hu}{}{Konkoly Observatory, MTA CSFK}{}{Budapest}{}{1121}{Hungary}
\paperauthor{\'A.~S\'odor}{sodor@konkoly.hu}{}{Konkoly Observatory, MTA CSFK}{}{Budapest}{}{1121}{Hungary}

\begin{abstract}
Previous studies of GD 244 revealed seven pulsation frequencies (two doublets and three single periods) in the light 
variations of the star. The data obtained at McDonald Observatory between 2003 and 2006, and our additional 
measurements in 2006 and 2007 at Konkoly Observatory, allow the investigation of the long-term pulsational behaviour 
of GD 244. We found that the 307.1\,s period component of one of the doublets show long-term, periodic amplitude modulation 
with a time scale of $\sim 740$ days. Possible explanations are that nonlinear resonant mode coupling is operating
among the rotationally split frequency components, or two modes, unresolved in the yearly data are excited at $\sim 307.1$\,s. 
This is the first time that such long-term periodic amplitude modulation is published on a ZZ Ceti star.
\end{abstract}

\section{Introduction}
GD 244 (V394 Peg, WD 2254+126, $V = 15.6$) was discovered to be a luminosity-variable DA white dwarf by 
\citet{f01}. Based on their 3-hour-long photometric observation with the Canada-France-Hawaii 
Telescope (CFHT), \citet{f01} determined four independent pulsation frequencies between $\sim 200-300$\,s 
and a few combination terms. According to its atmospheric parameters ($T_\mathrm{eff} = 11\,680$\,K, $\log g = 8.08$), 
GD 244 is situated in the middle of the empirical DAV instability strip. 
Photometric data were collected again on GD 244, between 2003 and 2006 (41 nights), with the 2.1-m Otto Struve 
Telescope at McDonald Observatory \citep{y05, m08}. The star was one of the targets of a survey, in which isolated 
and stable-amplitude modes were utilized to search for possible low-mass companions of pulsating white dwarf stars
applying the $O-C$ technique. The survey did not reveal any sign of a stellar- or planetary-mass companion of GD 244.

We collected data on GD 244 on 19 nights in 2006 and 2007 with the 1-m Ritchey-Chr\'etien-Coud\'e telescope at Piszk\'estet\H o 
mountain station of Konkoly Observatory. Fergal Mullally kindly provided the survey measurements on GD 244 for our analysis. 
These data and our additional measurements at Konkoly Observatory allow the detailed investigation of the long-term 
pulsational behaviour of GD 244. We present our first findings here. 

\section{Pulsation periods}

Table~\ref{tabl:periods} lists all the independent pulsation frequencies and the corresponding periods and amplitudes of GD 244 
published so far. Previous light-variation studies revealed seven pulsation frequencies, two doublets and three single 
periods. One of them (at 203\,s) proved to be considerably stable, and this was utilized to search for a possible 
planetary-mass companion in the survey of \citet{m08}. Not all the periods were detected 
in both data sets: e.g. the 294.6\,s mode found by \citet{f01} was not reported in the 2003 McDonald data 
\citep{y05}. There is a well-known general trend observed in the DAV stars that towards the red edge of the 
instability strip the light variations are more complex, non-sinusoidal, with longer periods and higher amplitudes. 
Short-term amplitude variations are also more common among the cooler ZZ Ceti stars (see e.g. \citealt{h03}), which may be 
responsible for the different set of periods observed in different epochs.

\begin{table}[!ht]
\caption{\label{tabl:periods}Pulsation frequencies of GD 244.}
\smallskip
\begin{center}
{\small
\begin{tabular}{lrrrr}  
\tableline
\noalign{\smallskip}
 & \multicolumn{3}{c}{McDonald Obs. data 2003$^a$} & CFHT data 1999$^b$\\
 & Freq. ($\mu$Hz) & Period (s) & Ampl. (mma) & Period (s)\\
\noalign{\smallskip}
\tableline
\noalign{\smallskip}
$f_1$ & 4926.7 & 203.0 & 4.0 & 203.3\\
$f_2$ & 3903.3 & 256.2 & 6.7 & 256.3\\
$f_3$ & 3897.7 & 256.6 & 12.3 & \\
$f_4$ & 	&	&	& 294.6\\
$f_5$ & 3261.9 & 306.6 & 5.0 & \\
$f_6$ & 3255.9 & 307.1 & 20.2 & 307.0\\
$f_7$ & 1103.7 & 906.1 & 1.7 & \\
\noalign{\smallskip}
\tableline 
\noalign{\smallskip}
\multicolumn{5}{l}{$^a$ \citet{y05}, $^b$ \citet{f01}}\\
\tableline\
\end{tabular}
}
\end{center}
\end{table}

\section{Amplitude modulation}

We checked the amplitudes and frequencies on daily, monthly and yearly time bases. We used \textsc{Period04} \citep{lb05},
and for comparison, custom developed software tools, including the light curve fitting program \textsc{LCfit} \citep{s12}. 
\textsc{LCfit} has linear (amplitudes and phases) and nonlinear (amplitudes, phases and frequencies) least-squares fitting options, 
and can handle base frequencies and linear combination terms as input.

Periods around 203, 256 and 307\,s can be detected practically in all daily data sets. The daily data suggested long-term 
periodic variation in the case of the 307\,s peak (see panel A of Fig.~\ref{fig:amplvar}), with $\sim 740$ day period. 
From one year to the next, the increase or decrease in the average of the daily amplitudes found to be 2-6 times larger 
than the corresponding standard deviations. However, we know that there are actually at least two peaks at 
307\,s, which cannot be resolved by the daily measurements, as their separation is $\sim 0.5$\,d$^{-1}$. Thus, we checked the 
amplitudes of the doublet components separately on longer time bases. We found that the 307.1\,s period component of the 307\,s 
doublet show clear long-term amplitude modulation. Neither the other component of this doublet (at 306.6\,s) nor any of the 
other known doublet's components (at 256\,s, see panel B of Fig.~\ref{fig:amplvar}) show similar variations. The 203\,s mode 
also remains stable both in amplitude (panel C of Fig.~\ref{fig:amplvar}) and in frequency. The frequencies of the doublet 
at 307\,s do not seem to vary (see Fig.~\ref{fig:freqs}).

\articlefigure[width=8cm]{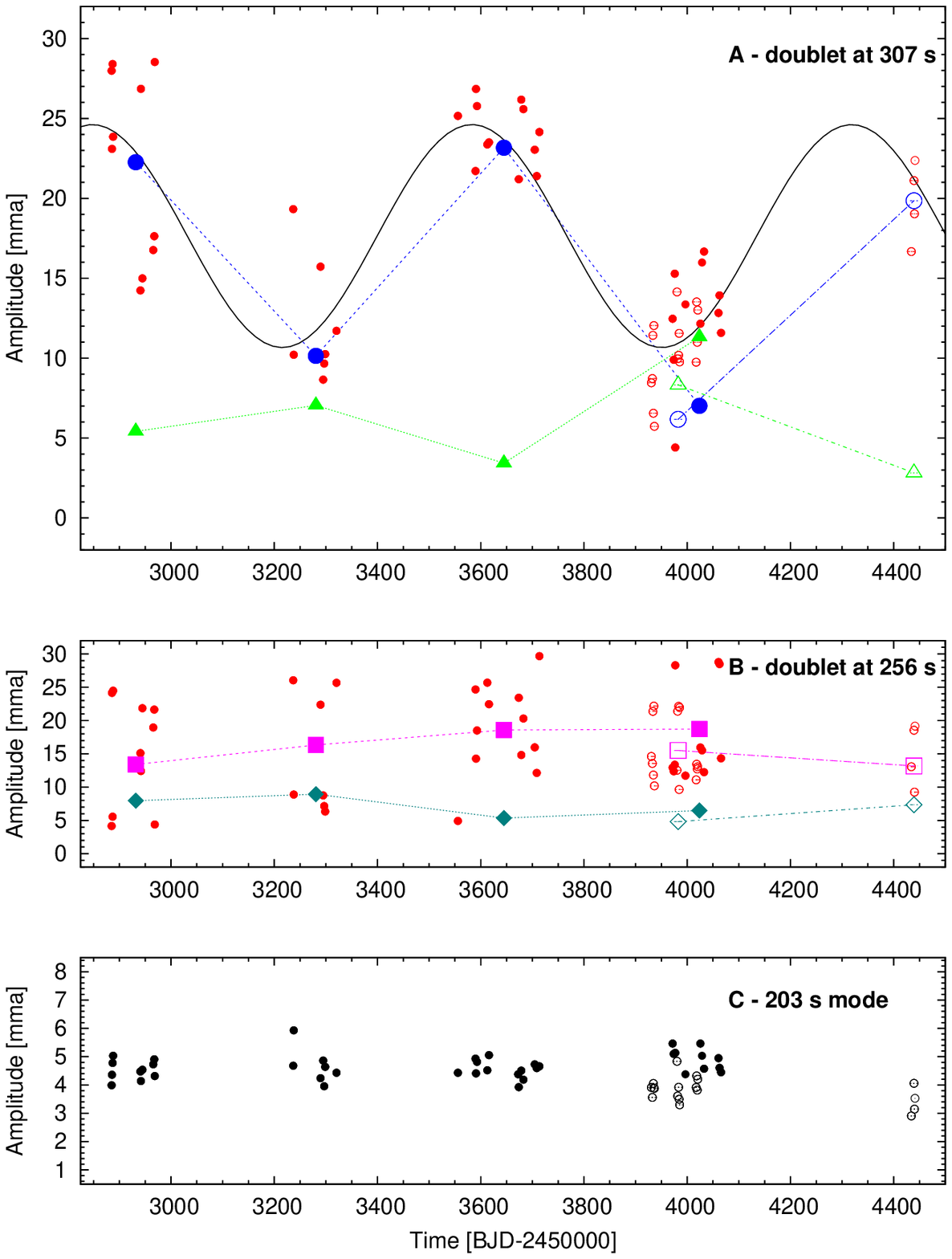}{fig:amplvar}{Amplitudes of different frequency components.
\textit{Filled} and \textit{open circles}: McDonald and Konkoly Observatory data, respectively.
\textit{Small red} (panels A and B) and \textit{black} (panel C) \textit{dots}: daily values.
\textit{Large blue dots} and \textit{green triangles} (panel A), \textit{magenta squares} and \textit{teal diamonds} (panel B): 
yearly amplitudes of the 307.1, 306.6, 256.6 and 256.2\,s peaks, respectively.
The standard errors of the fits are smaller than the sizes of the symbols used.
\textit{Black solid line} (panel A): sine wave fit to the daily data. The frequency used for the fit was fixed according to 
the frequency separation of the two closely spaced peaks found at $f_6$ by the combined 2003--2006 McDonald Observatory data
($\delta f = 0.00136$\,d$^{-1}, P \simeq 735$\,d, see Sect.~\ref{sect:disc}).
}

\articlefigure[width=8cm]{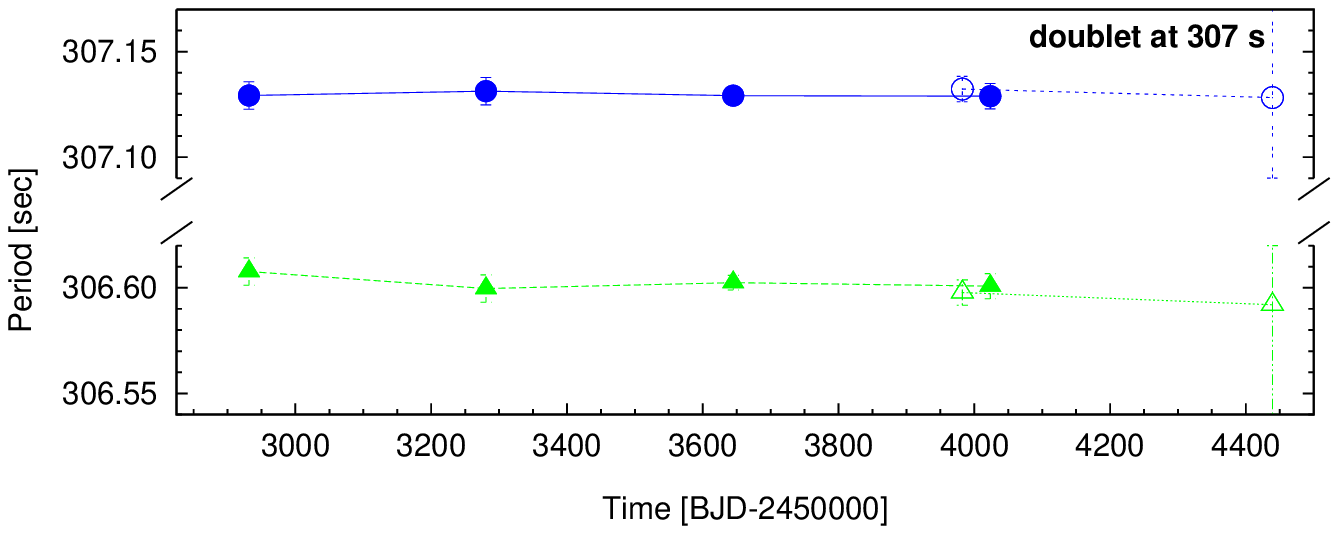}{fig:freqs}{Yearly periods of the doublet components at 307\,s. The errors are calculated as 
$\pm 0.5(1/\Delta\mathrm{T})$, that is, half of the Rayleigh frequency resolution.}

\section{Discussion and conclusion}
\label{sect:disc}

One possible explanation for the observed amplitude modulation is that two modes of fairly close frequencies are excited 
at $\approx 307.1$\,s. We analysed together all the 2003--2006 McDonald Observatory data for this purpose. For this preliminary 
analysis, we did not include the Konkoly data, to eliminate the possible instrumental effects caused by the different detectors 
and filters used for the observations.
The analysis revealed that we can detect indeed two significant peaks in this region of the Fourier spectrum of the combined 
yearly data, with the frequency separation of $\delta f = 0.00136$\,d$^{-1}$ (0.016\,$\mu$Hz). This is about three orders of 
magnitude smaller than the separation of the known $f_5$--$f_6$ doublet components. If these two peaks represent independent 
pulsation frequencies, they cannot be resolved by one-season-long observations, which can cause the observed apparent amplitude 
variation of the $307.1$\,s peak showed in Fig.~\ref{fig:amplvar}. This explanation also implies that both of these 
close-frequency modes have stable frequencies and amplitudes. Panel A of Fig.~\ref{fig:amplvar} shows a sine wave fit to the 
daily amplitude solutions with the frequency fixed to the $0.00136$\,d$^{-1}$ ($P \simeq 735$\,d) value.

Frequency and amplitude modulation in a similar time scale were reported in the case of the DBV-type \textit{Kepler} target 
KIC 08626021 \citep{z14}, which were explained as the results of nonlinear resonant coupling operating among 
rotational triplet components. Our case seems to be different: only one frequency component of the doublets shows clear 
sign of amplitude modulation, and we cannot find evidence of frequency modulation either. However, this cannot rule out
that this same physical reason is responsible for the observed variation, as nonlinear resonant coupling can cause
the amplitude modulation of only one frequency component. In this case, the closely spaced peak with $\delta f$
separation is only a mathematical representation of the amplitude modulation in the Fourier spectrum.

This is the first time that such long-term periodic amplitude modulation was published on a ZZ Ceti star. 

\acknowledgements \'A.S. acknowledges support by the J\'anos Bolyai Research Scholarship of the Hungarian Academy of Sciences.


\begin{thebibliography}{}
\bibitem[Fontaine et al.(2001)]{f01}
Fontaine, G., Bergeron, P., Brassard, P., Bill{\`e}res, M., \& Charpinet, S. 2001, ApJ, 557, 792
\bibitem[Handler(2003)]{h03}
Handler, G. 2003, in ASP Conf. Ser. Vol. 292, Interplay of Periodic, Cyclic and Stochastic Variability in Selected Areas of 
the HR Diagram, ed. C. Sterken (Astronomical Society of the Pacific: San Francisco), 247
\bibitem[Lenz \& Breger(2005)]{lb05}
Lenz, P., Breger, M. 2005, CoAst, 146, 53
\bibitem[Mullally et al.(2008)]{m08}
Mullally, F., Winget, D.~E., De Gennaro, S., Jeffery, E., Thompson, S.~E., Chandler, D., \& Kepler, S.~O. 2008, ApJ, 676, 573
\bibitem[S\'odor(2012)]{s12}
S\'odor, \'A. 2012, Konkoly Obs. Occ. Techn. Notes 15, 1\\ 
{\footnotesize \url{http://konkoly.hu/staff/sodor/lcfit.html}}
\bibitem[Yeates et al.(2005)]{y05}
Yeates, C.~M., Clemens, J.~C., Thompson, S.~E., \& Mullally, F. 2005, ApJ, 635, 1239
\bibitem[Zong et al.(2014)]{z14}
Zong, W., Charpinet, S., \& Vauclair, G. 2014, in this issue
\end{thebibliography}


\end{document}